\documentclass[journal,comsoc]{IEEEtran}

\usepackage[T1]{fontenc}
\usepackage{amsmath}
\usepackage{graphicx}

\usepackage{algorithm}
\usepackage{algpseudocode}
\usepackage{multicol}
\usepackage{float}
\usepackage{bm}
\usepackage{pifont}
\usepackage{relsize}
\usepackage[cmintegrals]{newtxmath}
\usepackage{mathrsfs}
\usepackage{subcaption}
\usepackage{multirow}

\ifCLASSINFOpdf
\else
\fi

\usepackage{amsmath}
\usepackage[cmintegrals]{newtxmath}
\usepackage{amsmath}
\usepackage{graphicx}

\usepackage{algpseudocode}

\begin{document}

\title{Non-orthogonal Multiple Access for Multi-cell Indoor VLC}

\author{T. Uday,~\IEEEmembership{Graduate Student Member,~IEEE}, Abhinav Kumar,~\IEEEmembership{Senior Member,~IEEE},
      and L. Natarajan,~\IEEEmembership{Member,~IEEE}
\thanks{The authors are with the Department of Electrical Engineering, Indian Institute of Technology Hyderabad, Telangana, India.
(e-mail: \{ee18resch11005, abhinavkumar, lakshminatarajan\}@iith.ac.in). This work was supported in part by the Department of Science and Technology (DST), Govt. of India (Ref. No. TMD/CERI/BEE/2016/059(G)).}}

\maketitle

\begin{abstract}
In this letter, we propose a non-orthogonal multiple access (NOMA) based scheme for multi-cell indoor visible light communications (VLC), where cell-edge user is jointly served by multiple cells. 
Unlike the typical designs in the existing literature, the proposed scheme doesn't involve complex domain and direct current bias addition, which makes the design simple and feasible for real time implementation in indoor VLC.
The symbol error rate (SER) of the proposed NOMA scheme is analysed using analytical and simulation results, and is compared with orthogonal multiple access (OMA). It is observed that the average SER of the users with the proposed NOMA is marginally degraded as compared to OMA. However, the SER of cell-edge user with the proposed NOMA is significantly improved with joint maximum likelihood decoding, and outperforms successive interference cancellation based decoding and OMA, with trade-off on computational complexity.

\end{abstract}

\begin{IEEEkeywords}
Cell-edge user, non-orthogonal multiple access, symbol error rate, visible light communications.
\end{IEEEkeywords}

\IEEEpeerreviewmaketitle
\section{Introduction}
\IEEEPARstart{V}{isible} light communications (VLC) is considered as an upcoming alternative to radio frequency (RF) communications especially in indoor scenarios where light emitting diodes (LEDs) are jointly used for illumination and communications \cite{intro_one}. Recently, non-orthogonal multiple access (NOMA) has been proposed for VLC \cite{intro_two}-\cite{intro_five}. In power domain NOMA, multiple users' signals are superposed in power domain at the transmitter and at the receivers, successive interference cancellation (SIC) based decoding is employed to retrieve the users' data. Most of the literature in NOMA-VLC considers single cell scenarios \cite{intro_two}-\cite{intro_five} where typically, orthogonal frequency division multiplexing (OFDM) based schemes have been considered. 
Several OFDM based schemes have been proposed for VLC such as direct current (DC)-biased optical OFDM (DCO-OFDM) and asymmetrically clipped optical OFDM (ACO-OFDM) \cite{intro_aco_dco_ofdm}. However, these techniques require Hermitian symmetry and inverse fast Fourier transform (IFFT) to convert the complex modulation symbols to real domain values. Then, DC biasing or clipping the negative part of the signal is required for DCO-OFDM and ACO-OFDM, respectively. The application of Hermitian symmetry reduces the sum rate of the users by half compared to radio frequency NOMA (RF-NOMA). 
\begin{figure}[!t]
\centering
\includegraphics[width=3.5in,height=1.35in]{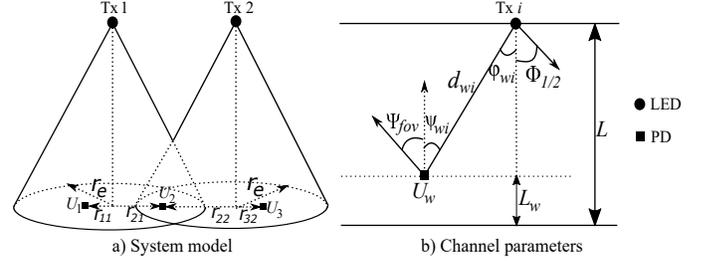}
\caption{System model and VLC channel parameters.}
\label{system_model}
\end{figure}
There have been a few studies on the NOMA for multi-cell scenarios \cite{intro_six}-\cite{intro_eight}. In \cite{intro_six}, a multi-cell scenario with three users, where one of the users is jointly served by both the cells has been considered. However, modulation scheme for such a scenario and theoretical analysis on error rate has not been discussed. A location based user grouping has been proposed in \cite{intro_seven} to reduce interference for multi-cell VLC networks. However, joint transmission for cell-edge users has not been discussed.  In \cite{intro_eight}, it has been shown that the joint transmission for the cell-edge users results in improved sum rate, but error performance of the users has not been presented. In \cite{self}, we have proposed a low complexity design for single cell NOMA-VLC system. However, multi-cell scenario has not been considered.
Therefore, in this paper, we design a low complexity and practically implementable scheme for  multi-cell indoor VLC scenario where cell-edge user is jointly served by multiple cells. The proposed scheme avoids DC bias or clipping and complex domain thereby avoiding Hermitian symmetry constraints which makes it easy for real time implementation in indoor environment.

The remainder of the paper is organized as follows. In Section II, we discuss the system model considered in this work. The proposed scheme is presented in Section III. In Section IV, we present the numerical results for the proposed scheme and comparisons from the literature. Some concluding remarks are discussed in Section V.
\section{System Model}
We consider a multi-cell scenario as shown in Fig.~\ref{system_model} (a), where the user in cell-edge is jointly served by the adjacent cells covering the user. Without loss of generality, we consider two cells each with a single transmitter (Tx) and two users, where, each Tx is equipped with a single LED and each user is equipped with a single photo diode (PD) at the receiver (Rx). One of the users is a cell-edge user (user with lower channel gain) common to both the cells and is jointly served by the Txs of both the cells using NOMA scheme proposed in the next section. The DC channel gain between $i^{th}$ Tx and $w^{th}$ user denoted by $h_{wi}$ is given as follows
$$
h_{wi}=
\begin{cases}
\frac{(\zeta+1)A_DR_p \textrm{cos}(\phi_{wi})^\zeta T(\psi_{wi})g(\psi_{wi})\textrm{cos}(\psi_{wi})}{2\pi d_{wi}^2}, \\
\,\,\,\,\,\,\,\,\,\,\,\,\,\,\,\,\,\,\,\,\,\,\,\,\,\,\,\,\,\,\,\,\,\,\,\,\,\,\,\,\,\,\,\,\,\,\,\,\,\,\,\,\,\,\,\,\,\,\,\,\,\,\,\,\,\,\,\,\,\,\,\,\,\,\,\,\,\,\,\,\, \psi_{wi}\in [0, \,\, \psi_{fov}],\\
0,\,\,\,\,\,\,\,\,\,\,\,\,\,\,\,\,\,\,\,\,\,\,\,\,\,\,\,\,\,\,\,\,\,\,\,\,\,\,\,\,\,\,\,\,\,\,\,\,\,\,\,\,\,\,\,\,\,\,\,\,\,\,\,\,\,\,\,\,\,\,\,\,\,\,\,\,\,\,\,\,\,\,\,\,\,\, \psi_{wi}>\psi_{fov}\,,
\end{cases}
$$
where $\zeta$ is the order of Lambertian radiation pattern given by $\zeta=-1/\textrm{log}_2(\textrm{cos}(\Phi_{1/2}))$  such that $\Phi_{1/2}$ is the angle at half power of LED, $A_D$ denotes the detection area of the PD at the Rx, $R_p$ denotes the responsivity of the PD, $T(\psi_{wi})$ represents the gain of the optical filter used at the Rx, where $\psi_{wi}$ is the angle of incidence at the $w^{th}$ user PD from $i^{th}$ Tx as shown in Fig.~\ref{system_model} (b), $d_{wi}$ is the distance between the $w^{th}$ user PD and the $i^{th}$ Tx, $g(\psi_{wi})$ represents the gain of the optical concentrator, $\psi_{fov}$ is the field of view of the PD at the Rx as shown in Fig.~\ref{system_model} (b), $\phi_{wi}$ angle of emission at the $i^{th}$ Tx with respect to (w.r.t.) $w^{th}$ user PD, $r_{wi}$ is the distance between Tx and PD from top view, $L$ is the height of the room, and $L_w$ is the position of the $w^{th}$ user's PD w.r.t. the floor of the room. In Fig.~\ref{system_model} (a), $r_e$ is the cell radius, $r_{11}$ is the top view distance between Tx1 and $U_1$, $r_{21}$ is the top view distance between Tx1 and $U_2$, $r_{22}$ is the top view distance between Tx2 and $U_2$, and $r_{32}$ is the top view distance between Tx2 and $U_3$.
\section{Proposed Scheme}
The encoding and decoding mechanism of the proposed scheme is as follows
\subsubsection{Encoding}
Let $P_{i,w}^{u_w}$ be the power allocated to $w^{th}$ user (denoted by $U_w$) by $i^{th}$ Tx, i.e., Cell $i$. Here, $u_w \in \mathcal{U}_w=\{1,\,2,\,\ldots,2^{\eta_w}\}\,\forall\, w \in \mathcal{W}= \{1,\,2,\,3\}$, $i \in \mathcal{I} = \{1,2\}$, and $\eta_w$ is the spectral efficiency of $w^{th}$ user in bits per channel use (bpcu).
 Similar to the work in \cite{self}, the allocated powers to the users are their respective constellation points.  Let the channel gains follow the order; $h_{11}>h_{21}$ and $h_{32}>h_{22}$, in Cell 1 and Cell 2, respectively, where $h_{11}$, $h_{21}$, $h_{32}$, $h_{22}$ are channel gain of $U_1$ in Cell 1, $U_2$ in Cell 1, $U_3$ in Cell 2, and $U_2$ in Cell 2, respectively.
Given this, the condition for zero BER in noiseless channel and perfect channel state information (CSI) conditions is given as follows
\begin{eqnarray}
\nonumber
P_{1,2}^{u_2} h_{21} + P_{2,2}^{u_2} h_{22} +
 \mbox{max}_{u_1}\big\{ P_{1,1}^{u_1}h_{21} \big\} +
 \mbox{max}_{u_3}\big\{ P_{2,3}^{u_3}h_{22} \big\} < \\
  \frac{\left(P_{1,2}^{u_2+1} h_{21}+P_{22}^{u_2+1}h_{22}\right) +\left(P_{1,2}^{u_2}h_{21} +P_{2,2}^{u_2}h_{22}\right) }{2}.
  \label{ber_constr_1}
\end{eqnarray}
We assume the constellation points of the users follows the following order
\begin{equation}
P_{i,w}^{1}<P_{i,w}^{2}<\ldots<P_{i,w}^{2^{\eta_{w}}}, s.t., |P_{i,w}^{u_w}-P_{i, w}^{u_w+1}|=\lambda_w,
\label{order_of_constellation}
\end{equation}
where $\lambda_w$ is the distance between consecutive constellation points of $w^{th}$ user.
In the received signal at $U_2$, the power allocated to $U_1$ and $U_3$ is treated as noise. In \eqref{ber_constr_1}, the maximum possible value of the received power from $U_1$ and $U_3$, i.e., $ \mbox{max}_{u_1}\big\{ P_{1,1}^{u_1}h_{21} \big\}$ and $\mbox{max}_{u_3}\big\{ P_{2,3}^{u_3}h_{22} \big\}$, respectively, is considered since they are the peak noise terms possible. For a given transmitted constellation point of $U_2$, the total received power at $U_2$ with zero channel noise should be less than the mid-point of the transmitted and it's consecutive constellation point of $U_2$ as given in the right hand side of \eqref{ber_constr_1}, to be successfully decoded.
From \eqref{order_of_constellation}, we have
$
\mbox{max}_{u_1}\big\{ P_{1,1}^{u_1}h_{21} \big\} = P_{1,1}^{2^{\eta_{1}}}h_{21}
$
and
$
\mbox{max}_{u_3}\big\{ P_{2,3}^{u_3}h_{22} \big\} = P_{2,3}^{2^{\eta_{3}}}h_{22}.
$
Given this, \eqref{ber_constr_1} can be simplified as follows
\begin{equation}
\frac{P_{1,2}^{u_2} h_{21}+P_{2,2}^{u_2} h_{22}}{2}+ P_{1,1}^{2^{\eta_{1}}} h_{21}+ P_{2,3}^{2^{\eta_{3}}} h_{22} <
\frac{P_{1,2}^{u_2+1} h_{21}+P_{2,2}^{u_2+1} h_{22}}{2},
\label{ber_constr_2}
\end{equation}
On further simplification of \eqref{ber_constr_2}, we get
\begin{algorithm}[t]
  \caption{Generate the constellation points for users.}
  \label{algo_convnoma}
  \begin{algorithmic}[1]
  \State Choose desired spectral efficiency for the users $\eta_w$, $w \in \{1,\,2,\,3 \}$ and $P$.
  \State Using \eqref{order_of_constellation} and assuming $\lambda_1=\lambda_3=1$, assign integer values for the constellation points corresponding to $1^{st}$  and $3^{rd}$ users (cell center users) as
  $1, 2, \ldots, 2^{\eta_w}$ for $w \in \{1,\,2 \}$. 
  
             \State $P_{1,2}^{1} = 2^{\eta_{1}}+1$ and $P_{2,2}^{1} = 2^{\eta_{3}}+1$.
                  \For{\texttt{$\mathit{u_2}=1$ to $2^{\eta_2}-1$}}
                  \State Find $P_{1,2}^{u_2+1}$ and $P_{2,2}^{u_2+1}$ such that
\begin{align*}
|\Delta_{U_2}^{u_2+1}  - \Delta_{U_2}^{u_2}-  \Delta| = 0,\\ \nonumber
P_{1,2}^{u_2+1} >  P_{1,2}^{u_2},\\ \nonumber
P_{2,2}^{u_2+1} >  P_{2,2}^{u_2}. 
\end{align*}
\State $P_{1, 2}^{u_2+1}=P_{1,2}^{u_2+1} +1$ and $P_{2,2}^{u_2+1}=P_{2,2}^{u_2+1} +1$. 
            \EndFor 
\State Normalize the constellation points using \eqref{norm_const_1} and \eqref{norm_const_2}.
\end{algorithmic}
\end{algorithm}
$
\Delta_{U_2}^{u_2+1}  > \Delta_{U_2}^{u_2}+  \Delta,
$
where $\Delta_{U_2}^{u_2+1}=P_{1,2}^{u_2+1} h_{21}+P_{2,2}^{u_2+1} h_{22}$, $\Delta_{U_2}^{u_2} =P_{1,2}^{u_2} h_{21}+P_{2,2}^{u_2} h_{22}$, and $\Delta=2 P_{1,1}^{2^{\eta_{1}}} h_{21}+ 2 P_{2,3}^{2^{\eta_{3}}}h_{22}$.
For transmission, the constellation points of the users in a given cell are normalized as follows
\begin{equation}
\widetilde{P_{1,w}^{u_w}} =  \left(\frac{ 2^{\eta_1+\eta_2} P}{\sum_{u_1=1}^{2^{\eta_1}}\sum_{u_2=1}^{2^{\eta_2}}\left( P_{1,1}^{u_1} + P_{1,2}^{u_2}\right)}\right)  P_{1,w}^{u_w},
\label{norm_const_1}
\end{equation}
$\forall\,u_w \,\in\,\mathcal{U}_w\,w\,\in\,\{1,2\}$, and
\begin{equation}
\widetilde{P_{2,w}^{u_w}} =  \left(\frac{ 2^{\eta_2+\eta_3} P}{\sum_{u_3=1}^{2^{\eta_3}}\sum_{u_2=1}^{2^{\eta_2}} \left(  P_{2, 3}^{u_3} +  P_{2,2}^{u_2}\right)}\right)  P_{2,w}^{u_w},
\label{norm_const_2}
\end{equation}
$\forall\,u_w \,\in\,\mathcal{U}_w\,w\,\in\,\{2,3\}$, where  '$\sim$' represents the normalized constellation point and $P$ is the desired average transmit power per channel use in any given cell.
Let $P_{1}$ and $P_{2}$ denote the peak transmit powers required by Tx1 and Tx2, respectively, to achieve the considered $P$. Given this,
\begin{equation}
\widetilde{P_{1,1}^{u_1}} +\widetilde{P_{1,2}^{u_2}} \leq P_{1},  
\label{mc_constraint_1}
\end{equation}
and
\begin{equation}
\widetilde{P_{2,3}^{u_3}} + \widetilde{P_{2,2}^{u_2}} \leq P_{2}. 
\label{mc_constraint_2} 
\end{equation}
In \eqref{mc_constraint_1} and \eqref{mc_constraint_2}, we consider $u_w=2^{\eta_w}\,\forall\,w\,\in\,\mathcal{W}$, since the maximum value of the constellation points assigned to $w^{th}$ user is for this value of $u_w$ from \eqref{order_of_constellation}. By this consideration, the transmit power required in any channel use and in any cell will not exceed the peak transmit power.
Given this, the algorithm to generate constellation points for the users is presented in Algorithm~\ref{algo_convnoma}.
Assuming additive white Gaussian noise (AWGN) channel as in \cite{intro_four}, the received signal at $U_1$, $U_2$, and $U_3$ denoted by $y_1$, $y_2$, and $y_3$, respectively, is given as follows
\begin{equation*}
y_1 = \left( \widetilde{P_{1,1}^{u_1}} + \widetilde{P_{1,2}^{u_2}} \right) h_{11} + n_1, 
\label{rx_sig_u1}
\end{equation*}
\begin{equation*}
y_2 = \sum_{w=1}^2 \widetilde{P_{1,w}^{u_w}}  h_{21} + \sum_{w=2}^3 \widetilde{P_{2,w}^{u_w}} h_{22} + n_2, 
\label{rx_sig_u2}
\end{equation*}
and
\begin{equation*}
y_3 = \left( \widetilde{P_{2,2}^{u_2}} + \widetilde{P_{2,3}^{u_3}} \right) h_{32} + n_3. 
\label{rx_sig_u3}
\end{equation*}
Here, $n_w$ is the AWGN noise at the $w^{th}$ user with zero mean and $\sigma_w^2$ variance and is denoted as $n_w \sim\mathcal{N}(0, \sigma_w^2)$. 
\subsubsection{Decoding}
The decoding at $U_1$ and $U_3$ needs SIC of $U_2$ as more power is allocated to $U_2$ because $h_{11}>h_{21}$ and $h_{32}>h_{22}$. Hence, the detection rule for the users is given as follows
\begin{equation*}
\widehat{ P_{1,1}^{u_1}}= \mbox{min}_{\widetilde{P_{1,1}^{u_1}}\forall u_1 \in \mathcal{U}_1}\left|\left|y_1 -h_{11}\widehat{ P_{1,2}^{u_2}} -h_{11} \widetilde{P_{1,1}^{u_1}}\right|\right|,
\end{equation*} 
where, $\widehat{ P_{1,1}^{u_1}}$ is the signal estimate of $U_1$, $\widehat{ P_{1,2}^{u_2}}$ is the signal estimate of $U_2$ SIC process at $U_1$ given by $\widehat{ P_{1, 2}^{u_2}}= \mbox{min}_{\widetilde{P_{1, 2}^{u_2}}\forall u_2 \in \mathcal{U}_2}\left|\left|y_1 -h_{11} \widetilde{P_{1,2}^{u_2}}\right|\right|$, and $||.||$ is the Frobenius norm. Similarly,
\begin{equation*}
\widehat{ P_{2,3}^{u_3}}= \mbox{min}_{\widetilde{P_{2,3}^{u_3}}\forall u_3 \in \mathcal{U}_3}\left|\left|y_3-h_{32} \widehat{ P_{2,2}^{u_2}} -h_{32} \widetilde{P_{2,3}^{u_3}}\right|\right|,
\end{equation*} 
where, $\widehat{ P_{2, 3}^{u_3}}$ is the signal estimate of $U_3$, $\widehat{ P_{2, 2}^{u_2}}$ is the signal estimate of $U_2$ SIC process at $U_3$ given by $\widehat{ P_{2,2}^{u_2}}= \mbox{min}_{\widetilde{P_{2,2}^{u_2}}\forall u_2 \in \mathcal{U}_2}\left|\left|y_3 - h_{32} \widetilde{P_{2, 2}^{u_2}}\right|\right|$.
Since $U_2$ has the highest allocated power, there are no SIC processes involved as the received power from lower power allocated users is treated as noise. Given this, the signal estimate at $U_2$, denoted by  
$\widehat{ P_{2}^{u_2}}$ is given as follows
\begin{equation*}
\widehat{ P_{2}^{u_2}}= \mbox{min}_{\widetilde{P_{2,2}^{u_2}} \forall u_2 \in \mathcal{U}_2}\left|\left|y_2-h_{21} \widetilde{P_{1,2}^{u_2}}-h_{22} \widetilde{P_{2,2}^{u_2}}\right|\right|.
\end{equation*}  
The lower power allocated users can cause performance degradation when the received power from them is treated as noise. Therefore, to improve the symbol error rate (SER) performance of $U_2$, we consider joint maximum likelihood (JML) decoding, where the ML decoding is performed jointly across all combinations of the constellation points of all users. The JML detection rule at $U_2$ is given as follows
\begin{equation*}
\widehat{ P_{2}^{u_2}}= \mbox{min}
\left|\left|y_2-h_{21}\left( \widetilde{P_{1,2}^{u_2}}+\widetilde{P_{1,1}^{u_1}}\right)
-h_{22} \left(\widetilde{P_{2,2}^{u_2}}+\widetilde{P_{2,3}^{u_3}}\right)\right|\right|,
\end{equation*}  
$ \forall \,\widetilde{P_{i,w}^{u_w}},\, \mbox{where,}\, \,u_w \in \mathcal{U}_w, w \in \mathcal{W},\, \mbox{and}\, i \in \mathcal{I} $.
Note that at $U_1$ and $U_3$, though JML decoding is possible, it is not employed since they do not have any corresponding lower power allocated users.
Considering the minimum distance decoding computations involved in SIC process and ML computations involved in ML decoding, 
the total number of computations involved in decoding process at $U_1$ and $U_3$ is $2^{\eta_1}+2^{\eta_2}$ and $2^{\eta_2}+2^{\eta_3}$, respectively. For $U_2$, the total number of computations using SIC based decoding is $2^{\eta_2}$ whereas using JML decoding, they are $2^{\sum_{w=1}^3\eta_w}$.
\subsubsection{Symbol error rate (SER)}
Now, we derive the expressions for SER of the users by assuming that the error propagation through SIC is zero.
\begin{table}[t]
\renewcommand{\arraystretch}{1.3}
\begin{center}
		\caption{Parameters for simulation \cite{intro_four}.}
		\label{sim_params}\begin{tabular}{ c|c }
\hline
Parameter   & Value \\\hline\hline
Cell radius, $r_e$ & $3.6\, m$\\
Height of the room, $L$ &  $4\,m$ \\
LED semi angle, $\Phi_{1/2}$ & $60^\circ$ \\
Optical filter gain, $T$ &  1 \\
PD FOV, $\psi_{fov}$ & $60^\circ$ \\
PD responsivity, $R_p$ & 0.4 $A/W$ \\
PD detection area, $A_D$ & $10^{-4} \,\, m^2$ \\
Position of $U_1$ PD w.r.t. floor, $L_1$ &  $0.5\,m$ \\
Position of $U_2$ PD w.r.t. floor, $L_2$ &  $0.5\,m$ \\
Position of $U_3$ PD w.r.t. floor, $L_3$ &  $1.0\,m$ \\
Refractive index of optical concentrator at Rx, $\eta$& 1.5\\
Top view distance between Tx1 and $U_1$, $r_{11}$ & $0.4885\,m$ \\
Top view distance between Tx1 and $U_2$, $r_{21}$ & $3.2880\,m$ \\
Top view distance between Tx2 and $U_2$, $r_{22}$ & $3.4670\,m$ \\
Top view distance between Tx2 and $U_3$, $r_{32}$ & $0.3030\,m$ \\\hline
\end{tabular} 
\end{center}
\end{table}

\textbf{Theorem 1:}
\textit{The SER of $2^{nd}$ user denoted by $P_{SER_2}$, using the SIC based decoding is given as follows}

\begin{align}\nonumber
P_{SER_2} = \left( 1-\frac{1}{2^{\eta_2}}\right)\frac{1}{2^{\eta_1+\eta_3}}\sum_{u_1=1}^{2^{\eta_1}}
\sum_{u_3=1}^{2^{\eta_3}} \\
\Bigg\{ Q\left(\frac{\rho^+(u_1,u_3)}{\sigma_2} \right) +  Q\left(\frac{\rho^-(u_1,u_3)}{\sigma_2} \right) \Bigg\},
\label{ser_u2}
\end{align}
\textit{where}
\begin{align*}
\rho^+(u_1,u_3) &= \frac{1}{2}\left|\widetilde{ P_{1,2}^{2}}-\widetilde{ P_{1,2}^{1}}\right| h_{21}+
 \frac{1}{2} \left|\widetilde{ P_{2,2}^{2}}-\widetilde{ P_{2,2}^{1}}\right| h_{22}\\
 &
-\left(h_{21} \widetilde{ P_{1,1}^{u_1}} +h_{22} \widetilde{ P_{2,3}^{u_3}}\right),
\end{align*}
\begin{align*}
\rho^-(u_1,u_3) &= \frac{1}{2}\left|\widetilde{ P_{1,2}^{2}}-\widetilde{ P_{1, 2}^{1}}\right| h_{21} +\frac{1}{2}\left|\widetilde{ P_{2,2}^{2}}-\widetilde{ P_{2,2}^{1}}\right| h_{22}\\
&
+\left(h_{21} \widetilde{ P_{1,1}^{u_1}} + h_{22}\widetilde{ P_{2,3}^{u_3}}\right),
\end{align*}
and 
$Q(t)=\int_t^{\infty} 1/\sqrt{2\pi}\,e^{-y^2/2}\, dy$.\\
\textbf{Proof:} Please see the Appendix for the proof. $\hfill\blacksquare$

\textbf{Remark 1:}
As the $U_1$ and $U_3$ do not have any noise terms from other users, the expression of SER is similar to the conventional pulse amplitude modulation (PAM) scheme and since we have assumed error propagation through SIC is zero, the SER expression can be lower bounded as follows
\begin{equation*}
P_{SER_1} > 2\left( 1-\frac{1}{2^{\eta_1}}\right)  Q\left(\frac{\left|\widetilde{ P_{1,1}^{u_1}}-\widetilde{ P_{1,1}^{u_1+1}}\right| h_{11}}{2\sigma_1} \right),
\label{ser_u1}
\end{equation*}
and
\begin{equation*}
P_{SER_3} > 2\left( 1-\frac{1}{2^{\eta_3}}\right)  Q\left(\frac{\left|\widetilde{ P_{2,3}^{u_3}}-\widetilde{ P_{2,3}^{u_3+1}}\right| h_{32}}{2\sigma_3} \right).
\label{ser_u3}
\end{equation*}
Note that the SER expression is an exact expression for $U_2$ since $U_2$ has the highest allocated power and thereby no SIC processes involved for error propagation through SIC. However, for $U_1$ and $U_3$, the SER expression is a lower bound since the error propagation through SIC decoding of $U_2$ at $U_1$ and $U_3$ is assumed to be zero.
\section{Numerical Results}
\begin{figure}[!t]
\centering
\includegraphics[width=3.5in,height=2in]{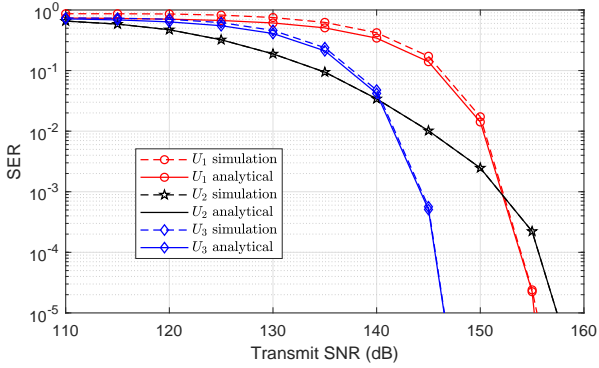}
\caption{SER of proposed NOMA showing simulation and analytical results for $\eta_1=3$ bpcu, $\eta_2=2$ bpcu, and $\eta_3=2$ bpcu.}
\label{ser_theo_sim}
\end{figure}
\begin{figure}[!t]
\centering
\includegraphics[width=3.5in,height=2in]{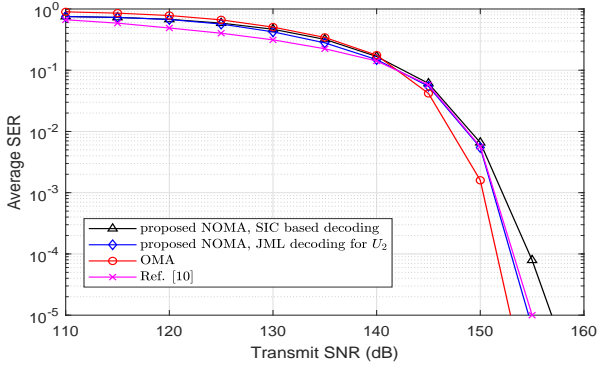}
\caption{Average SER performance comparison of proposed NOMA, OMA, and the scheme in [10].}
\label{ser_NOMA_Vs_OMA}
\end{figure}
The parameters used to plot numerical results are given in Table~\ref{sim_params}. Given this, the channel gains are computed as $h_{11}=2.5892 \times 10^{-6}$, $h_{21}=7.8573 \times 10^{-7}$, $h_{22}=6.8573 \times 10^{-7}$, and $h_{32}=3.5892 \times 10^{-6}$. 
The transmit SNR (dB) is defined as follows
\begin{equation*}
\mbox{Transmit SNR (dB)} = 10\, \mathrm{log}_{\mbox{10}}\left(\frac{P}{\sigma^2}\right)\,\mbox{dB},
\end{equation*}
where, $\sigma^2=\sigma_1^2=\sigma_2^2=\sigma_3^2$. Though the average transmit power for $U_2$ is twice compared to $U_1$ and $U_3$, the noise is from two channels which makes the transmit SNR of $U_2$ same as that of $U_1$ and $U_3$.
\begin{figure}[!t]
\centering
\includegraphics[width=3.5in,height=2in]{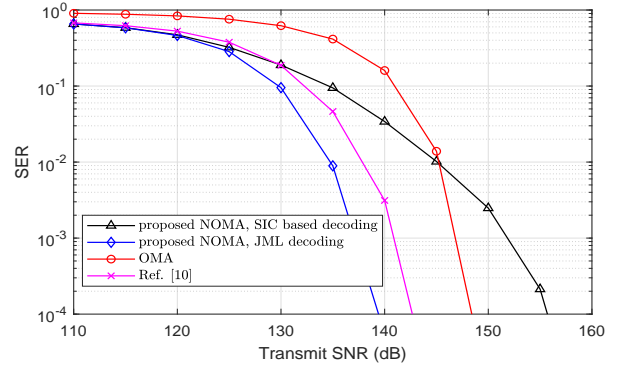}
\caption{SER of the cell-edge user, $U_2$ with proposed NOMA using SIC based decoding and JML decoding, OMA, and the scheme in [10]}
\label{ser_NOMA_Vs_OMA_U2}
\end{figure}
In Fig.~\ref{ser_theo_sim}, the SER of the proposed NOMA is plotted showing simulation and analytical results for $\eta_1=3$ bpcu, $\eta_2=2$ bpcu, and $\eta_3=2$ bpcu. It is observed that the simulation result exactly match with the analytical expression for $U_2$, whereas, the analytical expressions for $U_1$ and $U_3$ are close bounds as mentioned in Section-III (3). At high SNRs, the performance of cell-edge user, $U_2$ is deteriorating because $U_2$ has the highest allocated power and the power received from lower power allocated users is treated as noise while decoding at $U_2$.

We define average SER as the average of the SER of all the three users.
In Fig.~\ref{ser_NOMA_Vs_OMA}, the average SER performance of proposed NOMA with SIC based decoding and JML decoding for $U_2$ is compared with OMA and the scheme in \cite{self}. For fair comparison, the spectral efficiencies of the users are scaled for OMA such that $\eta_1=6$ bpcu, $\eta_2=4$ bpcu, and $\eta_3=4$ bpcu. The scaling ensures that for any given user, the number of bits transmitted per channel use is exactly same for proposed NOMA and OMA. In OMA, data can be simultaneously transmitted to $U_1$ and $U_3$ by Tx1 and Tx2, respectively, in the same channel use since $U_1$ and $U_3$ are in the non-overlapping cell regions. In the consecutive channel use, both Tx1 and Tx2 jointly transmits the data to $U_2$.
The constellation points of the proposed NOMA and OMA are normalized such that the average transmit power per channel use remains same for both the schemes. The constellation points of the $M$-ary PAM considered for OMA are generated using
\begin{equation*}
I_m=\frac{2I_{avg}m}{M+1}, \, m=1,\,2,\,\ldots,M,
\end{equation*}
where, $I_m$ is the $m^{th}$ constellation point and $I_{avg}$ is the average intensity level. 
Since the scheme proposed in \cite{self} is for a single cell scenario, we consider NOMA in Cell 1 by assuming $U_2$ to be in Cell 1 since $U_2$ has higher channel gain in Cell 1 than in Cell 2. Given this, $U_3$ is the only user in Cell 2, and hence, OMA is used in Cell 2. We use JML decoding for $U_2$ in Cell 1. Since $U_2$ in \cite{self} is assumed to be only in Cell 1, there will be interference from Cell 2 which we assume to be zero for simulations.
It is observed that though the average SER performance of proposed NOMA with SIC based decoding and JML decoding for $U_2$ is marginally poor compared to OMA. However, the cell-edge user's performance is improved considerably with the proposed NOMA using JML decoding, as shown in Fig.~\ref{ser_NOMA_Vs_OMA_U2}. It is also observed that the average SER of scheme in \cite{self} is as good as the proposed NOMA. However, the SER performance of cell-edge user in proposed NOMA is considerably better compared to the performance with the scheme in \cite{self}.
For the considered spectral efficiencies, the computational complexity of proposed NOMA with SIC based decoding and JML decoding for $U_2$, OMA and the scheme in \cite{self} is compared in Table~\ref{computational_complexity}. It is observed that SER gain with the JML decoding in proposed NOMA is achieved at the cost of increased computational complexity.

\begin{table}[t]
\renewcommand{\arraystretch}{1}
\begin{center}
		\caption{Comparison of the computational complexity.}
		\label{computational_complexity}\begin{tabular}{ |c|c|c| }
\hline
 \multirow{3}{*}{Scheme}   & Average number of & Computations  \\
 & computations & of cell-edge user\\
 & per channel use & per channel use\\\hline
 Proposed NOMA with &\multirow{2}{*}{24} &\multirow{2}{*}{4} \\
 SIC based decoding  & & \\\hline
  Proposed NOMA with &\multirow{2}{*}{148} & \multirow{2}{*}{128}\\
 JML decoding for $U_2$  & & \\\hline
 \multirow{2}{*}{OMA} &\multirow{2}{*}{48} & \multirow{2}{*}{8}\\
  & & \\\hline
 Scheme in \cite{self} with &\multirow{2}{*}{48} &\multirow{2}{*}{32} \\
 JML decoding for $U_2$ & & \\\hline

\end{tabular} 
\end{center}
\end{table}
\section{Conclusions}
In this letter, we have proposed a NOMA based scheme for multi-cell indoor VLC, where cell-edge user is jointly served by multiple cells. The SER performance of the proposed NOMA scheme is analysed using analytical and simulation results, and is compared with OMA. It is observed that though the average SER performance of the users with the proposed NOMA is marginally degraded as compared to OMA and as good as the existing NOMA in \cite{self}, the SER of the cell-edge user is significantly improved with the proposed NOMA using JML decoding with trade-off on the computational complexity.
\section*{Appendix}
\subsection{Proof of Theorem 1:}
If there were no noise terms, the decision boundary between the consecutive constellation points would be their mid-point, denoted by $\gamma$ and is given as follows
\begin{equation}
\gamma = \frac{1}{2}\left|\widetilde{ P_{1,2}^{u_2+1}}-\widetilde{ P_{1, 2}^{u_2}}\right| h_{21}+ \frac{1}{2} \left|\widetilde{ P_{2,2}^{u_2+1}}-\widetilde{ P_{2, 2}^{u_2}}\right|h_{22}.
\label{proof_eq1}
\end{equation}
From \eqref{order_of_constellation}, since the distance between each pair of consecutive constellation points is same, \eqref{proof_eq1} can be written as follows
\begin{equation}
\gamma = \frac{1}{2}\left|\widetilde{ P_{1, 2}^{2}}-\widetilde{ P_{1,2}^{1}}\right| h_{21}+ \frac{1}{2} \left|\widetilde{ P_{2,2}^{2}}-\widetilde{ P_{2,2}^{1}}\right|h_{22}.
\label{proof_eq2}
\end{equation}
However, there will be a change in the actual boundary due
to the noise terms. Here, the noise terms are the constellation
points of lower power allocated users, i.e., $U_1$ and $U_3$. These noise terms will cause a shift in the decision boundary towards the next constellation point as these noise terms are always positive real and additive. Given this, and using \eqref{proof_eq2}, the decision boundary w.r.t. next constellation point, denoted by $\rho^{+}(u_1,u_3)$ is given as follows
\begin{align}\nonumber
\rho^+(u_1,u_3) = \gamma - \left(h_{21} \widetilde{ P_{1,1}^{u_1}} +h_{22} \widetilde{ P_{2,3}^{u_3}}\right),\\ \label{proof_eq3}
\forall\,u_1\in\{1,\,2,\,\ldots,2^{\eta_1} \}, \,u_3\in\{1,\,2,\,\ldots,2^{\eta_3} \}. 
\end{align}
Similarly, the decision boundary w.r.t. previous constellation point, denoted by $\rho^{-}(u_1,u_3)$ is given as follows
\begin{align}\label{proof_eq4}
\rho^-(u_1,u_3) = \gamma + \left(h_{21} \widetilde{ P_{1,1}^{u_1}} +h_{22} \widetilde{ P_{2,3}^{u_3}}\right).
\end{align}
Averaging the error over all the constellation points of $U_1$ and $U_3$ using \eqref{proof_eq2}, \eqref{proof_eq3}, and \eqref{proof_eq4}, the final expression for SER is given as in \eqref{ser_u2}.
This concludes the proof of Theorem 1.$\hfill\blacksquare$

\end{document}